\documentclass{elsart}

\usepackage{isolatin1}
\usepackage[dvips]{graphicx}
\usepackage{color}
\usepackage{amssymb}
\usepackage{amsmath}
\usepackage{epsf}
\usepackage{subfigure}
\begin{document}

\begin{frontmatter}

\title{Modeling viral coevolution: HIV multi-clonal persistence and competition dynamics}

\author{F. Bagnoli\thanksref{a}\thanksref{c}\thanksref{1}},
\author{P. Li\`{o}\thanksref{b}\thanksref{2}},
\author{L. Sguanci\thanksref{a}\thanksref{c}\thanksref{3}}

\address[a]{Dipartimento di Energetica, Universit\`{a} di Firenze, Via S. Marta 3, 50139 Firenze, Italy.}
\address[b]{Computer Laboratory, University of Cambridge, CB3 0FD Cambridge, UK.}
\address[c]{INFN, sezione di Firenze and CSDC.}

\thanks[1]{franco.bagnoli@unifi.it}
\thanks[2]{pietro.lio@cl.cam.ac.uk}
\thanks[3]{luca.sguanci@unifi.it}

\begin{abstract}
The coexistence of different viral strains (quasispecies)
within the same host are nowadays observed for a growing number
of viruses, most notably HIV, Marburg and Ebola, but the 
conditions for the formation and survival of new strains have not
yet been understood. We present a model of HIV quasispecies competition,
that describes the conditions of viral quasispecies coexistence under
different immune system conditions. Our model incorporates both T and
B cells responses, and we show that the role of B cells is important
and additive to that of T cells. 
Simulations of coinfection (simultaneous infection) and
superinfection (delayed secondary infection) scenarios
in the early stages (days) and in the late stages of the infection
(years) are in agreement with emerging molecular biology findings. 
The immune response induces a competition among similar phenotypes,
leading to differentiation (quasi-speciation), escape dynamics and
complex oscillations of viral strain abundance. 
We found that the quasispecies dynamics
after superinfection or coinfection has time scales of
several months and becomes even slower when the immune system 
response is weak. Our model represents a general framework to study 
the speed and distribution of HIV quasispecies during disease progression, 
vaccination and therapy.

\end{abstract}

\end{frontmatter}

\section{Introduction}
\label{sec:Introduction} The human immune system has the goal of
providing a basic defense against pathoge\-nic organisms
\cite{PC2001,PW1997}. The cells forming the immune system
communicate via direct contact and through
chemical signals. Using a very basic schematization, the immune
system protective action can be divided into an innate, aspecific
response, and an adaptive one.  This latter component  involves
two major types of cells: lymphocytes, (T and B lymphocyte cells)
and antigen presenting cells: monocytes, macrophages, B
cells and dendritic cells. Lymphocytes have the ability to react to
specific \emph{epitopes}, which  generally are portions of a
protein. The idea is to recognize those shapes that constitute a
fingerprint of the presence of a foreign agent (antigen), but not
to react to \emph{self} repertoire. Their response depends
on the T cell receptor (TCR) of the T cell, and immunoglobulins
(Ig) of the B cells. It is the great variety of cells
carrying TCRs or secerning Igs that allows the
recognition of all possible antigens.

Let us present a simplified description of the adaptive response,
see Ref.~\cite{PW1997} for a physicists-oriented introduction to
immune system modeling. The task of protecting the body implies
the detection and removal of free antigens, and the recognition
and suppression of infected (or cancer) cells. These
\emph{deviating} self cells are identified by a sophisticated
automatic mechanism, common to all cells, that presents on their
surface specific receptors along with epitopes. 
Both viral infection or cancer proliferation implies the
activation of \emph{novel} genes, and thus the production of
\emph{foreign} proteins that can thus be detected. However, due to
the danger (and cost) of an inflammation, the activation of the
immune adaptive response is subject to a \emph{double key}
protection, implemented by T helper cells.

T cells are divided into CD4+ (helper) and CD8+ (killer) types,
according with specific receptors present on their surface. CD4+ T
cells provide the regulatory mechanism that triggers the immune
response by finding and presenting antigens and releasing
stimulatory signals (cytokines). T cells are specific for a
given set of epitopes; simplifying, we can say that they are
specific for a given antigen. Once activated, CD8+ T cells search
for and kill other cells presenting the specific antigen.
Their action is triggered by CD4+ T cells recognizing that
antigen. This specificity is not absolute: there is a
certain variability of receptors allowing cross-recognition of similar
epitopes.

The coverage of all possible shapes is implemented by a unique mechanism of genetic  combinatorial sampling.
The protection against self recognition is due to the maturation phase.
Immature T lymphocytes are
generated in the bone marrow and then travel to the thymus where they
proceed through a selection process. In the thymus those CD4+ and CD8+ T
cells that initiate any reaction against \emph{self} shapes are selected out.

B cells are responsible of the removal of extracellular antigens.
Antigen specific CD4+ T cells are required to induce proliferation
of antigen specific B cells. The activated B cells then
proliferate and differentiate into plasma cells. These produce
soluble counterparts to the cell's receptors, called antibodies,
which circulate in the blood and fluids. Antibodies bind to
specific extracellular antigens and allow them to be removed by
other means.

An antigen selects a ``strain'' of lymphocytes which replicates in
response (clonal response). During this growth, the specificity of
receptors to a given antigen may be increased by an hypermutation
mechanism that slightly modifies the genes specifying the receptor
shapes. A few cells of this strain (memory cells) will persist for
a long time in the body, after viral eradication: should the
specific antigen reappear, the response will be faster and
stronger (vaccination).

Let's now describe the immune response in the framework of a HIV
infection. HIV infects cells that contain the CD4 receptor on
their cell membrane, through which the virus attaches and enters the cell. 
The CD4 molecules are found on T lymphocyte helper cells, 
monocytes and macrophages.

As HIV uses the CD4 molecule to establish the initial binding to a
cell, these cells are the main target for HIV infection. Since
CD4+ T cells are vital to the establishment of an immune response,
their progressive loss opens the way for AIDS-related
opportunistic infections~\cite{PF1993}.

The worldwide presence of several strains of the HIV virus and
their often simultaneous presence within a patient, due to the
increased frequency of multiple infections, are the new remarkable
features of HIV pandemia. For example, HIV-1, presently, exists as
three groups classified according to their degree of similarity.
Group M stands for main, Group O for outlier and Group N for
non-M, non-O. Group M is the predominant group and is divided into
9 strains (subtypes or strains) that differ in the env gene~\cite{KM2004,YL2005}. 
Specific strains are predominant in different
geographical regions and their high mutation rates often allow
researchers to track the network of contacts. A growing number of
strains within each subtype is nowadays reported.

The origin of different HIV strains is related to the quasispecies
characteristics of HIV evolution~\cite{ES1989}. Quasispecies are the
combined result of mutations and recombination,
originating variability, and selection, that filters out this variability. 
The main source of mutations is the reverse transcription
phase~\cite{PC2001}, in which the viral RNA is converted into DNA by viral
proteins that are not very accurate. Other sources of variability
are the errors during the transcription phase, deletions and other
genetic rearrangements during the integration of viral DNA  into
the host chromosomes, including recombination with other viral DNA
or proviruses present in the cell. This mechanism thus increases
the viral diversity in the body. However, not all these variants
have the same capability of completing their life cycle and
multiplying their clone number (i.e.\ their fitness).

First of all, the infection capacity of mutants may vary, and also
their speed of replication~\cite{NP1998}. In some cases a single mutation is
sufficient to alter significatively the fitness~\cite{KH1997}, or
this may due to the accumulation of several mutations (smooth or
\emph{Fujiama} landscape~\cite{FP1997}). Since the number of
targets (the substrate) is limited, the strains that originate
more clones in the unit of time tend to eliminate less fitness
mutants, which are subsequently regenerated by the mutation
mechanism~\cite{ES1977}. While mutations are an essential
ingredient for exploring the genetic space in the search for the
fitness maximum, they also lowers the average fitness of the
strain, that generally is formed by a cloud of mutants around the
fitness maximum (quasispecies). For a given fitness landscape,
there is a maximum tolerable mutation rate above which the
quasispecies structure is lost (error catastrophe~\cite{ES1989}).

The HIV quasispecies seem able to compete and their persistence
affect AIDS progression ~\cite{CD2000,BL2005}. For
instance, some mutants may even be unable to colonize new cells,
but they may survive by parasitizing more invasive clones (this is
particularly observed when recombination of several strains is
present)~\cite{JD2000}. 
Finally, the immune response is specific, with a
certain tolerance, to a given antigen, and may vary from antigen
to antigen, due, for instance, to their similarities to self
shapes. Thus, the presence of an antigen originates an immune
response that suppresses also variants of the stimulating clone.
This results in a form of competition among strains or clones,
that prefer to stay \emph{away} in the genetic space in order to
escape the immune response and/or lower their intensity by dividing
the productive capacity of the immune system into several strains
(multi-infection). The effect of this competition may result in a
continuously varying dominant quasispecies that tries to escape
the immune response, or the splitting of the quasispecies into
multiple strains (quasi-speciation).  During the first phases of
HIV infection, when the immune response is vigorous, the first
phenomenon is observed until presumably the virus finds a
\emph{niche} in the genetic space, not sufficiently covered by the
immune system, originating AIDS~\cite{DSP1992,HM1995}. The
resistance to antiviral treatment may be ascribed to a similar
mechanism. When the immune system is depressed, also slightly less
adapted strains may survive, giving origin to multi-clonal
infections often observed in terminal patients~\cite{BM2002}.

Secondary HIV infections with closely related strains may induce
the response and clonal selection of a new subsets of CD4+ T cells
that recognize specifically the new HIV strain, even under
conditions of chronic infection by other strains. We can
distinguish two scenarios: strains coinfection (simultaneous
infection) and superinfection (delayed secondary infection)
scenarios (see for instance Refs.~\cite{R2002,L2003,HH2004,AW2002,GH2005}).

Our aim is to model viral strain coevolution and understand the
importance of factors such as mutation rates, strength of immune
response, cross-talks between T and B cells and competition among
strains in coinfection and superinfection.

The fate of an epidemic does not depend only on the evolution
inside a single host, but also on the transmissibility to other
hosts. The social factors involved in emergence of viral strains
are diverse and include global transportation, urban crowding and
poverty, changing behavioral patterns, human population growth,
etc. \cite{ZH1996}. We shall not deal with these aspects in the
present paper.

\begin{figure}[t]
  \centering
  \includegraphics[width=0.6\textwidth]{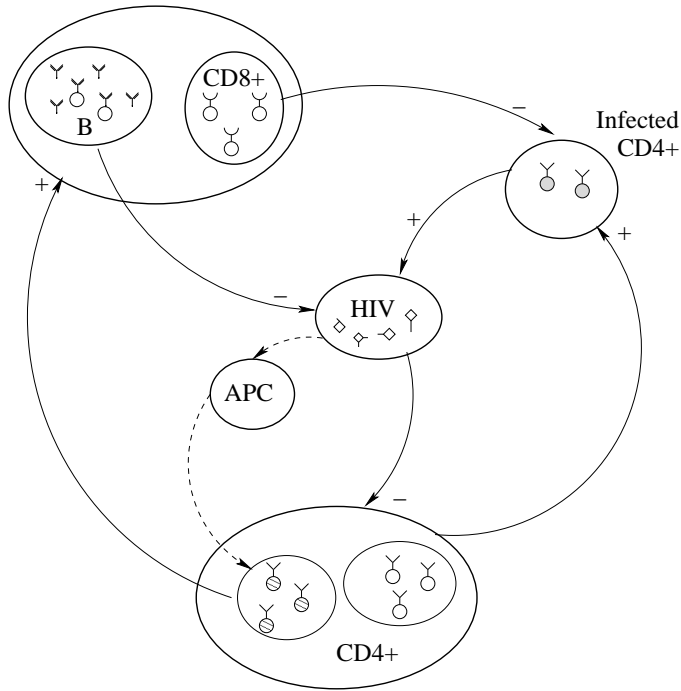}
  \caption{Representation of the interactions between cells and/or cells and
viruses. HIV strains 
infect CD4+ T cells that become infected and produce new viruses.
At the same time HIV peptides are presented via APC cells to
T-Helper cells that become activated. Activated CD4+ T cells
trigger B and CD8+ cells reactions: the first release antibodies
that bind to the antigen while the latter directly remove infected
CD4+ T cells. '+' and '-' signs indicate cell/virus production or
removal.}
  \label{fig1}
\end{figure}

The article is organized in the following way: in the next
sections we introduce a model that combines T and B cells immune
response to HIV infection. Then we extend the model to include
quasispecies and we test the model in the scenarios of
coinfection and superinfection using parameters derived from
biological literature. Finally we discuss results and work in
progress.

In what follows, we shall use the terms strain and quasispecies,
according with the \emph{scale} at which one is examining the system.
In order to resolve the quasispecies details, one needs to use a
(genotypic or phenotypic) scale sufficiently fine to allow
differentiation inside a single quasispecies. On the
contrary, if one is interested in the \emph{ecological} relations
among quasispecies, it is more efficient to use a scale in which a
whole quasispecies is simply collapsed in a single value. In this
latter case we use the term strain to refer to the whole quasispecies.

\section{Models}

Mathematical models have proven valuable in understanding the
mechanisms of many of the observed features of HIV dynamics,
such as the positive and
negative regulation of T and B cell selection, the dynamics of
production of the TCR repertoire in the thymus, the progression to
and the latent phase of
AIDS~\cite{HM1995,CP2004,CS1996,DP1995,WN2002,PH1996,WP2004,WS1995}.
They have also been useful in forecasting the effects of multi drug therapy.

While the majority of HIV dynamics
models assume that each CD4+ T cell is infected by a single HIV
strain, Dixit and Perelson~\cite{DP2005} have shown that with the
progression of HIV infection multiple infections become very
likely.

Our goal was to investigate the conditions of the
multi-strains persistence within a single patient
and the effects of coinfection and superinfection. To set the scenarios of our
model, it is noteworthy that, in general, the immune response can
be seen as a prey/predator dynamics, in that the immune system is
the predator of the virus (the prey). The HIV dynamics is more
complex because the HIV is both the prey and the predator, i.e. it
attacks CD4+ T cells and is attacked by cells coordinated by CD4+
T cells.

Figure~\ref{fig1} illustrates the relationships among the
different quantities of our model. This schematization is valid
for the quasispecies case and for the
undifferentiated one. In this latter case, the
indication of strains (in parenthesis) should be
neglected. The
clonal amplification of naive T cells (of population $i$) depends on
the ability of ($i^{\text{th}}$ class) T cell to recognize all the infected
T cells (carrying an epitope from the $k^{\text{th}}$ class of viruses).
The rate of infection of naive T cells (of class $i$) depends on all
the viruses (containing the epitope $k$). In the same way, the
clonal expansion of infected T cells (of class $k$, meaning that
they have been infected by a virus of class $k$ irrespective of the
original T class) depends on the
interaction of these viruses with all the T cells (of any class $i$).

\subsection{Combining B and T cells responses}

\begin{figure}[t]
  \centering
  \includegraphics[width=0.6\textwidth]{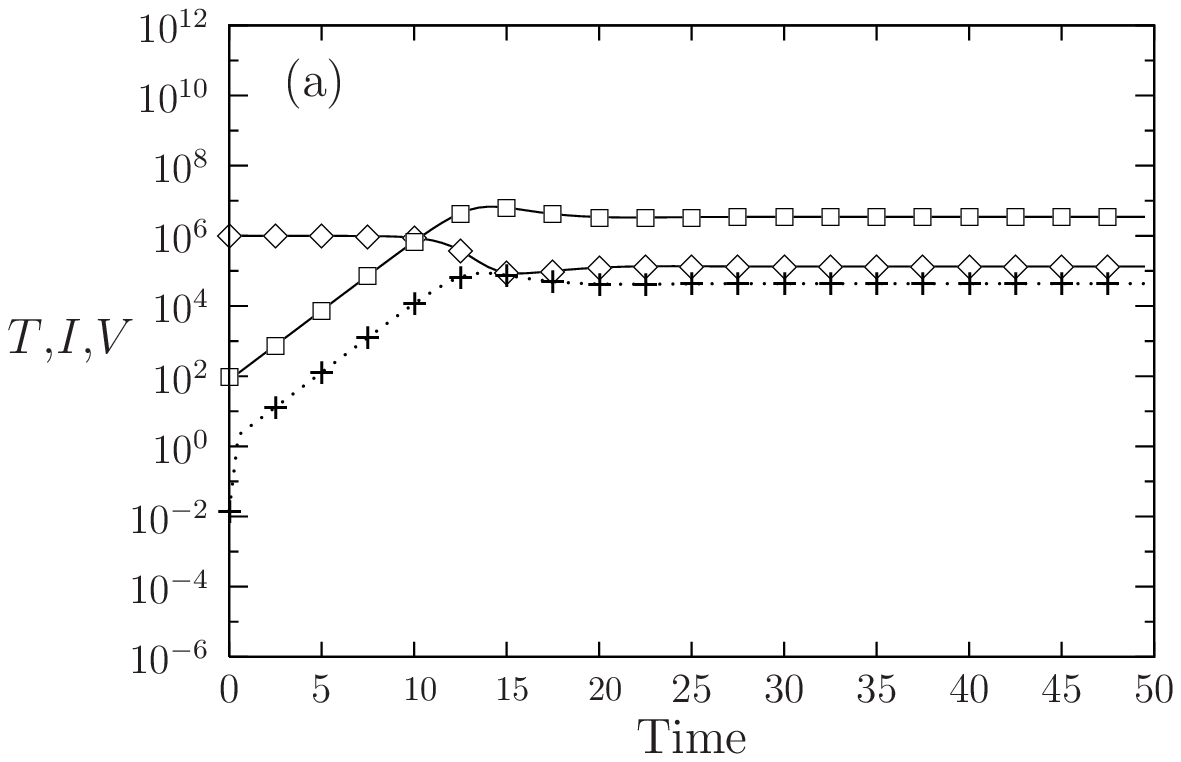}
  \includegraphics[width=0.6\textwidth]{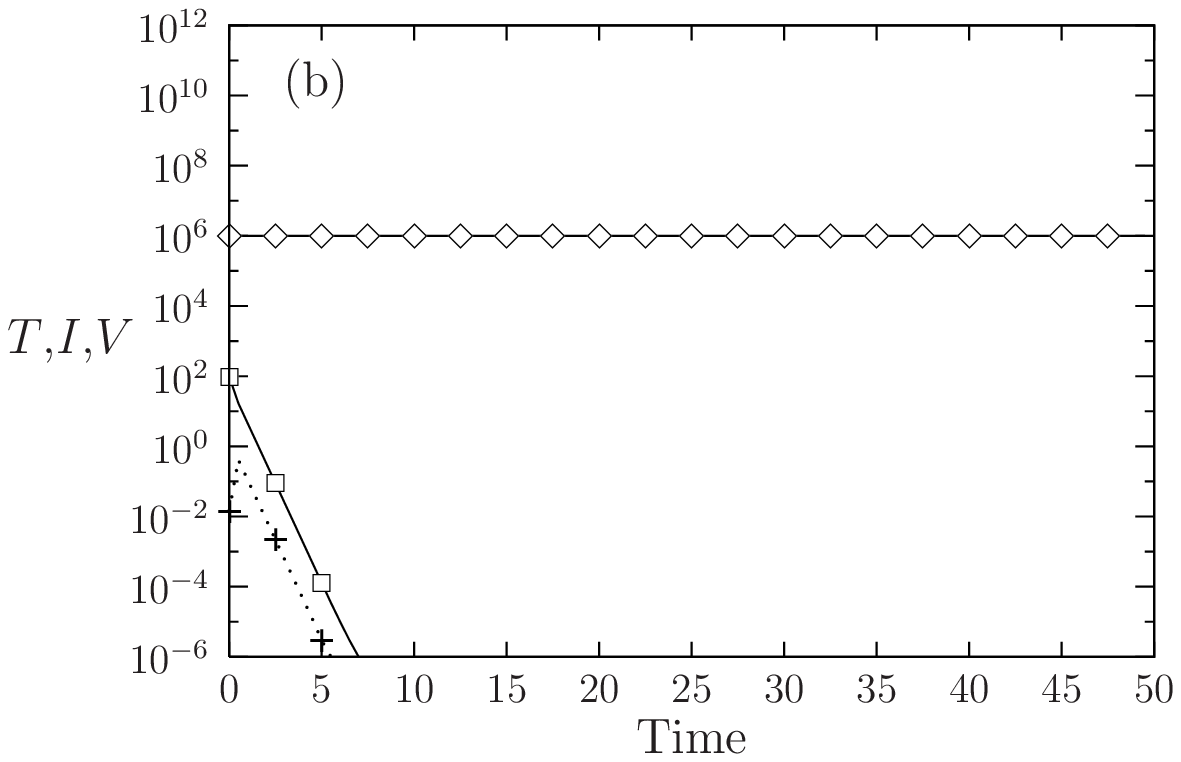}
  \caption{Typical time evolution of the single-species model; 
  diamonds represent
uninfected T cells, squares represent viruses and plus signs the
infected T cells. Plot (a) illustrates a
  scenario leading to a chronic infection
  ($\gamma^{(T)}=10^{-6}$, $\gamma^{(I)}=10^{-5}$, $\gamma^{(V)}=0$).
 In the presence of a further B cell response (plot (b)), the
infection is defeated by the immune system
($\gamma^{(T)}=10^{-6}$, $\gamma^{(I)}=10^{-5}$ and
$\gamma^{(V)}=5\cdot10^{-6}$).}
  \label{fig2}
\end{figure}

Let us start from a widely accepted theoretical model of HIV
progression and immune response proposed by Perelson and
colleagues in 1995~\cite{HM1995}, recently extended to include CTL
response and stochastic components~\cite{CP2004}. There are
several works that discuss the role of B cells in the immune
response~\cite{DSP1992}. Since the progression to AIDS has been
found to correlate well with CD4+ T cells decrease, B cells are
thought to play a minor role in the immune response to HIV. Note
that B cells can act only as predator to the HIV, so their
coupling with HIV dynamics is different from that of T cells.
Our aim is to present a more general  model framework of
both T and B immune responses to HIV. We have first considered the
following system of differential equations describing the dynamics
of a single viral strain:
\begin{align}
\dot{T}  &= \left(  \lambda+\gamma^{(T)}I\, T\right)
  \left(  1-T/K\right)-\left(  \delta_{T}+\beta\, V\right)  T, \label{T}\\
\dot{I}  &= \beta\, V\,T-\left(  \delta_{I}+\gamma^{(I)}\,T\right)
I,\label{I}\\
\dot{V}  &= \pi\, I-\left(  c+\gamma^{(V)}\,T\right)  V.\label{V}
\end{align}
All quantities indicated by greek letters and $c$ have unit
$\text{days}^{-1}$.

This model considers the T-helper (CD4+) cells ($T$), and HIV
virus particles ($V$); the T cells can become infected ($I$). With
respect to Refs.~\cite{HM1995,CP2004}, in Eq.~\eqref{T}, we
describe how the number of naive T cells ($T$) which have passed
the thymus selection, depends on rate of formation in the bone
marrow ($\lambda$), and on clonal amplification upon stimulation
by infected cells, $I$ (term $I\, T$). They decrease with a rate
that is the sum of a natural clearance, $\left( \delta_{T}T\right)$,
 due to cell aging, and cell destruction upon virus infection,
$\left( VT\right)  $. The density of T cells is limited by a
saturating density/lymphonode capacity factor, $K$. Following
Ref.~\cite{WP2004} we have set $K=10^{12}$.

The second equation
describes the rate of infection, described by $\beta$, of naive T
cells upon the interaction with the virus (term $V\,T$). Infected T cells
are cleared out at a fixed rate, $\delta_{I}$, and due to the
action of natural killer cells, CD8+ (term $TI$).

The third equation
describes the budding of viruses from infected cells, $\pi$. Virus
particles are cleared out at rate $c$ (defective viruses) and
after immunoglobulin binding and subsequent engulfments by the
macrophages
(term $TV$). 

The $\gamma$ parameters have the same meaning of the
constant of association in chemistry, or can be thought as a
combination of both the probability of interaction and the
interaction strengths between cells ($\gamma^{(T)}$,$\gamma^{(I)}$) or
between cells and viruses ($\gamma^{(V)}$). Note that in the limit
$\gamma \rightarrow 0$ and $K\rightarrow\infty$, 
we recover the pattern of the standard
model~\cite{PH1996}.

The B cell response is  modeled using the
parameters corresponding to the T cells which activate them by
receptor recognition. Here, we have assumed the immunoglobulin
concentration, which represent the B cell response, are linearly
correlated to the concentration of activated B cells. These  in
turn  are supposed linearly correlated to the concentration of the
CD4+ T cells. Exploratory analysis with different
parameters, suggested us that Eq.~\eqref{V} allows to keep a
minimum number of parameters without loss of important details.

Figure~\ref{fig2} shows some typical behavior of this single-species
model, showing the case of asymptotic coexistence of immune system and
virus (chronical infection) and virus eradication. This latter
scenario may be triggered both by increasing the T killer
($\gamma^{(I)}$) or B ($\gamma^{(V)}$) response. In no cases it is
possible to observe the defeat of the immune system, since a decrease
of the quantity of T cells, affects also the rate of viral production.
This reflects the observed fact that progression to AIDS and death is
caused by opportunistic infections and not directly by the HIV virus.
Model parameters are derived from medical
literature (see also Ref.~\cite{CP2004}).

\subsection{Modeling quasispecies competition and persistence}

Our actual model is obtained by extending Eqs~(\ref{T}-\ref{V}),
to include the effects of quasispecies persistence.

For the sake of simplicity, we
assume that each viral strain is characterized by just one epitope
$i$. We make use of coupled differential equations, one for each
viral quasispecies and T cell. Although this mean-field approach
disregards the effect of fluctuations and genetic drift in
quasispecies abundances, it is useful for understanding the
coarse-grain features of the behavior of the interplay between HIV and
the immune system.

The quasispecies model is described by the following set of equations:
\begin{align}
\dot{T}_{i}  &= \left(  \lambda_{i}+\sum_{k}\gamma_{ik}^{(T)}I_{k}T_{i}\right)
  \left(  1-\frac{1}{K}\sum_{i}T_{i}\right)  -
  \left(  \delta_{T}+\sum_{k}\beta_{k}V_{k}\right)T_{i}  ,\label{Ti}\\
\dot{I}_{k}  &= \left(  \sum_{k^{\prime}}\mu_{kk^{\prime}}
\beta_{k^{\prime}}V_{k^{\prime}}\right)\left(\sum_{i}T_{i}\right)
 -\left(  \delta_{I}+\sum_{i}\gamma_{ki}^{(I)}T_{i}\right)I_{k}
 ,\label{Ik}\\
\dot{V}_{k}  &= \pi
I_{k}-\left(c+\sum_{i}\gamma_{ki}^{(V)}T_{i}\right)V_{k}.\label{Vk}
\end{align}

The model considers the following cell types: T-helper (CD4+)
cells responding to virus strain $i$, ($T_{i}$); T cells (any strain)
infected by virus strain $k$, ($I_{k}$); abundance of viral strain
$k$, ($V_{k}$).
This means that viral strain
$k$ are identified by just one epitope, which is then displayed
on the surface of the
T cell of class $k$,  and that a T cell of class $i$ can be activated at
least by one CD4+ T cell carrying the epitope $k$, which is
specific of the viral strain $k$.
The indices $i$ ($k$) range from 1 to $N_i$ ($N_k$),
and in the following we have used $N_i=N_k=N$.

Equation~\eqref{Ti} describes the generation of T cells through two
mechanisms: the bone-marrow source (and selection in the thymus)
and the duplication of T cell strains activated upon the
recognition with an antigen carrying cell that may be even an
infected one. In order to take into account the limits of the
immune response, we have investigated both a logistic term and
different types of saturating and density-dependent functions~\cite{CP2004}.
The logistic term models the
global carrying capacity of immune system. In other
works~\cite{CP2004}, a saturating function or density dependent
function was used, aiming to
describe how the growth and development of the immune response is
limited by the timing of T cell division among other constraints
such as cell-cell recognition, signals diffusion, cell migration.

The death-rate term represents both a natural
death rate proportional to the population, and the infection rate of T
cells by any viral strain. The infection probability, reflected by
the term $\sum_{k}\beta_{k}V_{k}T_{i}$ and by the sum over $T_{i}$ in the
$I$ cell birth rate, is the same irrespective of the T class. We
assume that an infected T cell does no more contribute to the immune
response.

Equation~\eqref{Ik} describes the infection
dynamics. The incorporation of two death rate parameters reflects
the fact that the infected cells disappear due to cellular
death and after the action of natural killer cells (CD8+). There
are clear experimental evidences that CD4+ cells decrease during
the late HIV infection stages and in the AIDS state. In terms of
our models this mean that the parameter $\delta_{I}$
depends on the stage of the
infection. We have preferred to model the early stage of
infection, far from AIDS conditions and therefore we have set
$\delta_{I}$ to a constant, medical literature referred, value.

The term $\mu_{kk^{\prime}}$ represents a mutation process that
affect the phenotype. Mutations occur mainly during the reverse
transcriptase processing of the viral genomes. The mutation parameter 
is essential for the formation
of new quasispecies. Although RNA viruses have been reported to
have substitution rates of the order of $1 \cdot 10^{-3}$
substitution per site per replication~\cite{JH2002}.
Mansky and Temin~\cite{MT1995}
determined an in-vivo mutation rate of $3.5 \cdot 10^{-5}$.
The majority of mutations does not
change the amino acids. Moreover, most of amino acid substitutions are neutral
or quasi-neutral, since they do not change remarkably the protein
structure, and therefore the fitness of the species; effective
differences in fitness and behavior evolve through several amino
acid changes (see for instance \cite{LG2004}). Nevertheless, one
single event of recombination can often alter the fitness in a
substantial way. We take into consideration only those non-synonymous
mutations that alter the phenotype (protein structure),
and therefore we used a slight smaller
value for the mutation rate of the order of $10^{-5}$.

We have implemented the matrix $\mu_{kk^{\prime}}$ as a discrete
Laplacian (diffusion in phenotypic space) controlled by parameter
$\mu$ (without indices).

Equation~\eqref{Vk} describes the virus replicative dynamics in terms of a
birth rate proportional to the virus "budding" numerosity.
The viral death rate parameters depend on the  rate of natural death and on the
recognition of virus by B cells.

As in the single-specie model, B-cells and T-killer cells are only
implicitly included in the model in order to reduce the
dimensionality without loosing too many details. We assume that
these responses are fast enough to be at equilibrium and they are
just proportional to the abundance of (cognate) $T$ helper cells.

While in the single-strain model, Eqs.~(\ref{T}-\ref{V}), 
the three $\gamma$ parameters are scalar values,
here $\gamma_{ik}^{(T)}$, $\gamma_{ki}^{(I)}$ and
$\gamma_{ki}^{(V)}$ are matrices describing the interactions
between viral strains and immune cells, i.e.\ who will interact
with whom. The $\gamma$ parameters represent the interaction
between cells and/or cells and viruses in terms of geometry and
strength of the interaction. It expresses in a single matrix all
the information about diffusion, epitope recognition and show some
analogies with a constant-in-time fitness, as described in Ref.~\cite{BB1997}.

When the $\gamma$ matrices are diagonal, each T cell
interacts only with one viral strain. The non-zero
elements of the $i^{\text{th}}$ row in the  $\gamma$ matrices
represent the strains of the virus
recognized by the immune cells of class $i$ while the value represents
the affinity maturation, i.e.\ the accuracy in recognition.
 
\begin{figure}[t]
\centering
\includegraphics[width=0.5\textwidth]{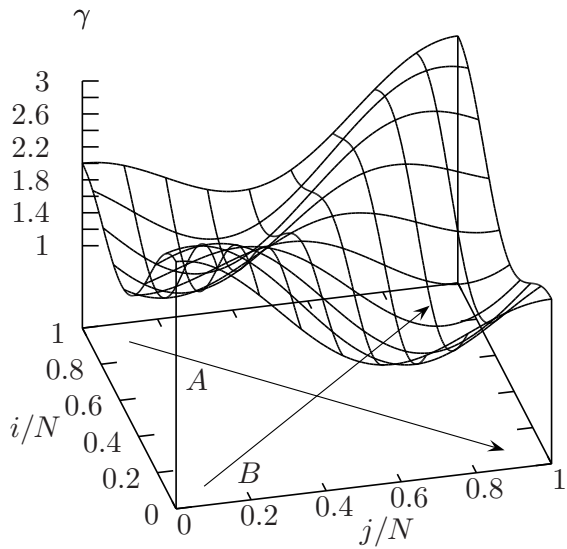} 
\caption{Control parameters and shape of the $\gamma$-matrices}
\label{gamma}
\end{figure}
 
In order to control the shape of the $\gamma$ matrices with few
parameters, we choose the following functional form (see
Fig.~\ref{gamma})
\[
  \gamma_{ij} = 
A\left\{\frac{1}{2}\left[1+\cos\left(\frac{2\pi}{N}(i-j)\right)\right]\right\}^{\varepsilon_A}
+
B\left\{\frac{1}{2}\left[1+\cos\left(\frac{\pi}{N}(i+j)\right)\right]\right\}^{\varepsilon_B}.
\]
The parameters $A$ and $\varepsilon_A$ ($B$ and $\varepsilon_B$) 
control the shape of the
$\gamma$ matrices in the direction transverse (parallel) to the main
diagonal. 
The exponents $\varepsilon_A$ and $\varepsilon_B$ controls the smoothness of the
variation along the corresponding direction.  For 
$\varepsilon$ small the variation of the function
is very small, while for $\varepsilon\rightarrow \infty$  the
corresponding function is $\delta$-shape. So, a diagonal $\gamma$
matrix is obtained by setting $B=0$ and $\varepsilon_A$ large. 
For instance, differences in the fitness  of phenotypes may be
obtained by setting $B$ and $\varepsilon_B$ different from zero in
some of the $\gamma$-matrices, while the intra-species competition is
triggered by setting $A$ different from zero and $\varepsilon_A$ small.

It is worth noting that nowadays, 2-photons experiments allow to
measures recognition strength, movements and speed of the immune
cells~\cite{MC2004} that, in theory, can be used to derive estimates of the
$\gamma$ matrices. We think that this data is still in embryonal phase,
so we focus on the qualitatively features of the response.

Given the absence of a precise relationship
between genotype and phenotype changes, we have simply considered
the grouping together all quasispecies with the same fitness. Thus
the quasispecies constitute a linear space ordered in term of
phenotype similarity.

We have used the same values of model parameters by re-scaling
$T,I,V$, $\lambda_{i}$ and $K$ by the number $N$ of strains.

\section{Numerical results}

\begin{figure}[t]
  \centering
  \includegraphics[width=0.4\textwidth]{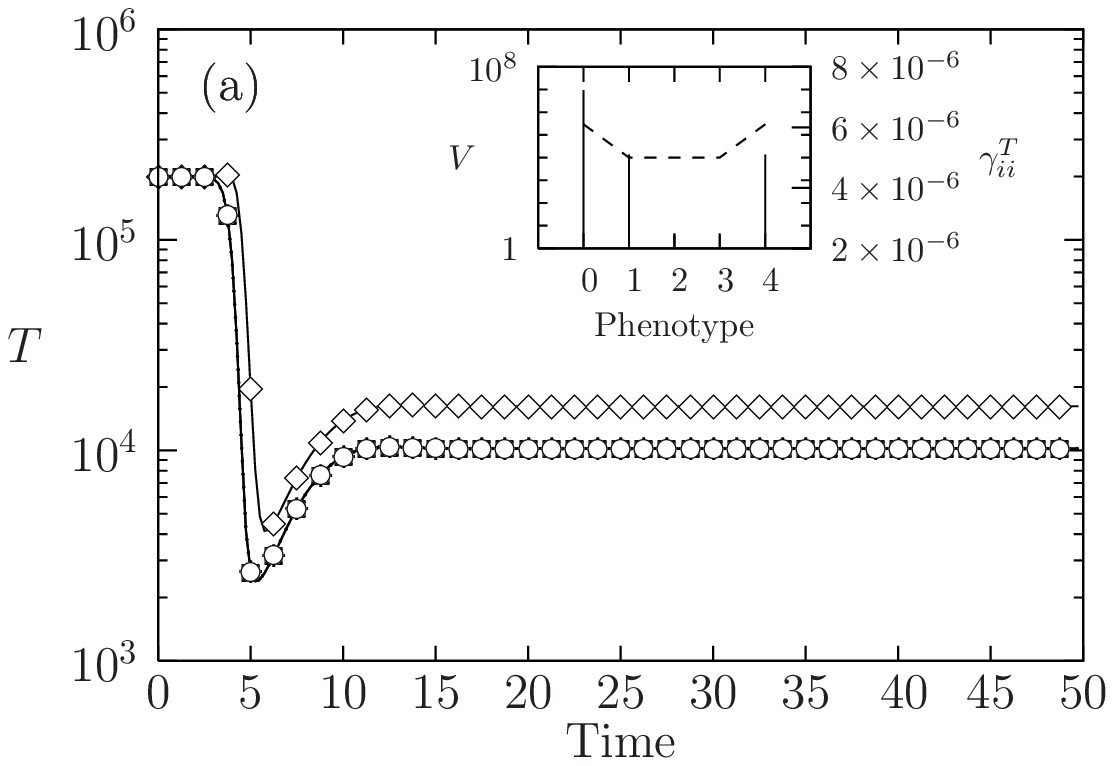}
  \includegraphics[width=0.4\textwidth]{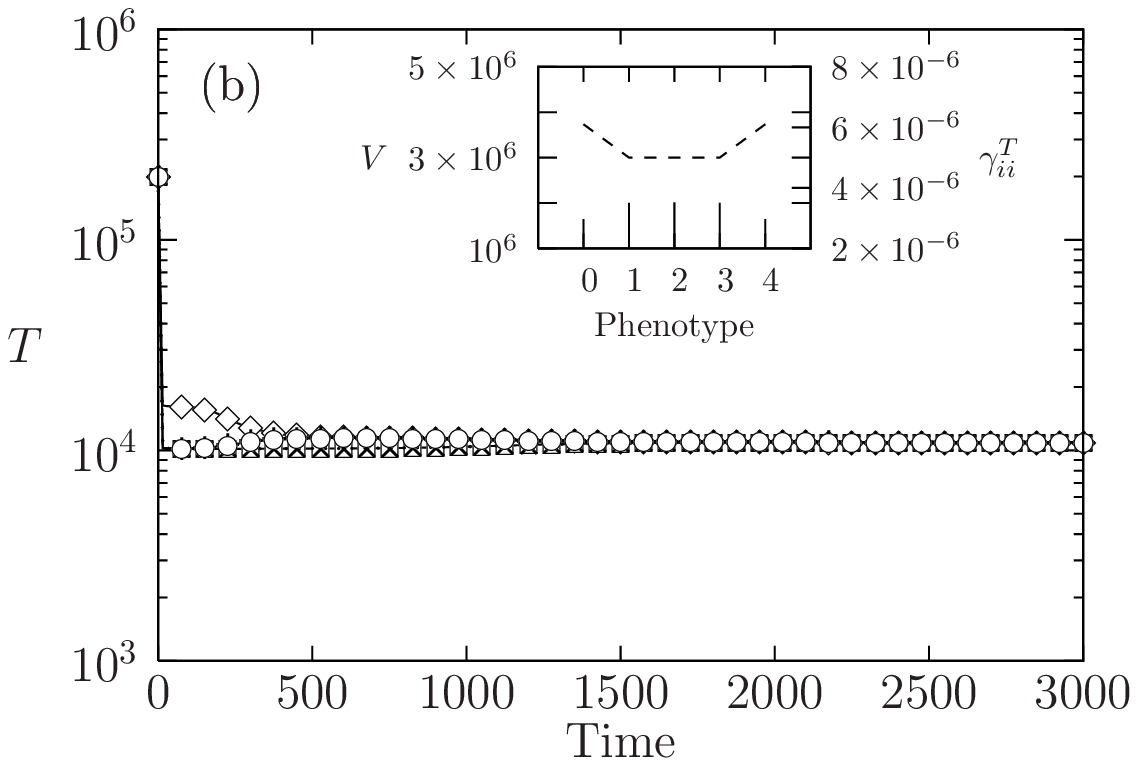}
  \caption{Typical time evolution of the T cell abundance
    in the multi-species model for short-term (a) and long-term
    behavior (b). We set $\mu=10^{-5}$,  
    $A = 5\cdot 10^{-6}$, $\varepsilon_A=10$, $B^{(T)} = 3\cdot
    10^{-6}$, $\varepsilon_B^{(T)}=10$ with $N=5$.
    In the inset, the y-axis on the left reports the asymptotic abundance of
    virus strains, the y-axis on the right shows the interaction
    strength (dashed line) between T cells and virus phenotypes (x-axis).}
\label{fig3}
\end{figure}

\begin{figure}[t]
  \centering
   \includegraphics[width=0.4\textwidth]{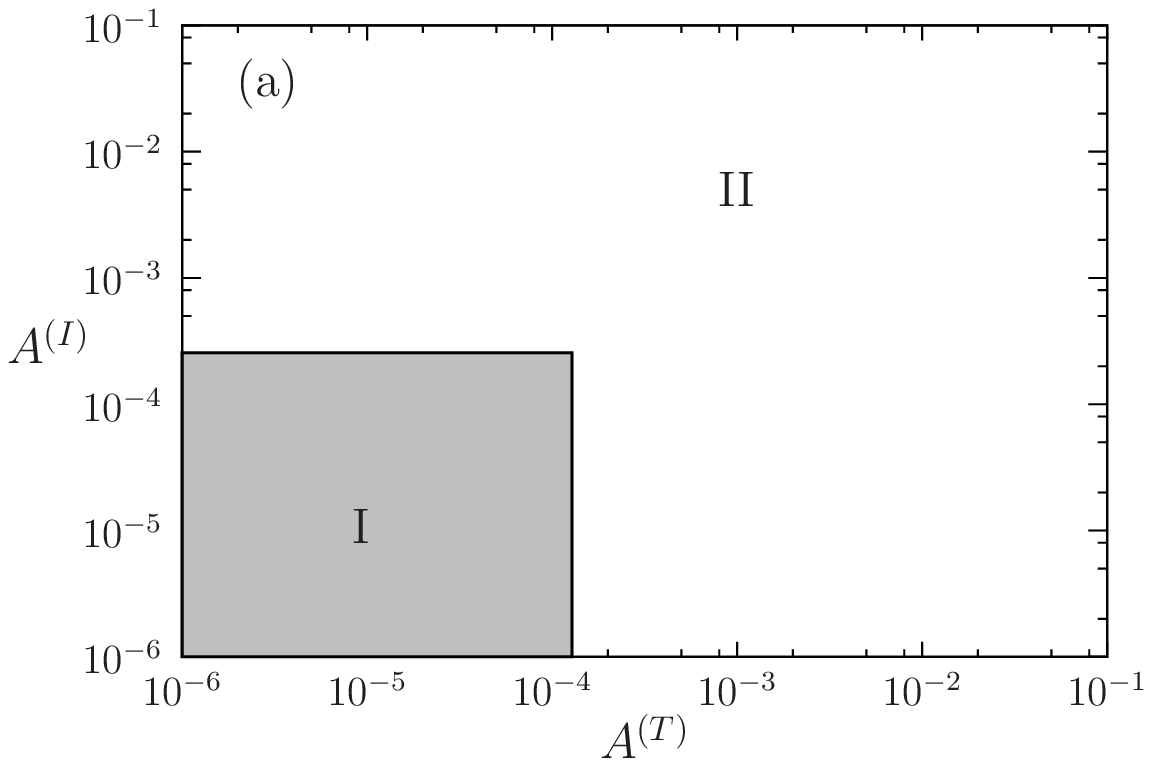} 
    \includegraphics[width=0.4\textwidth]{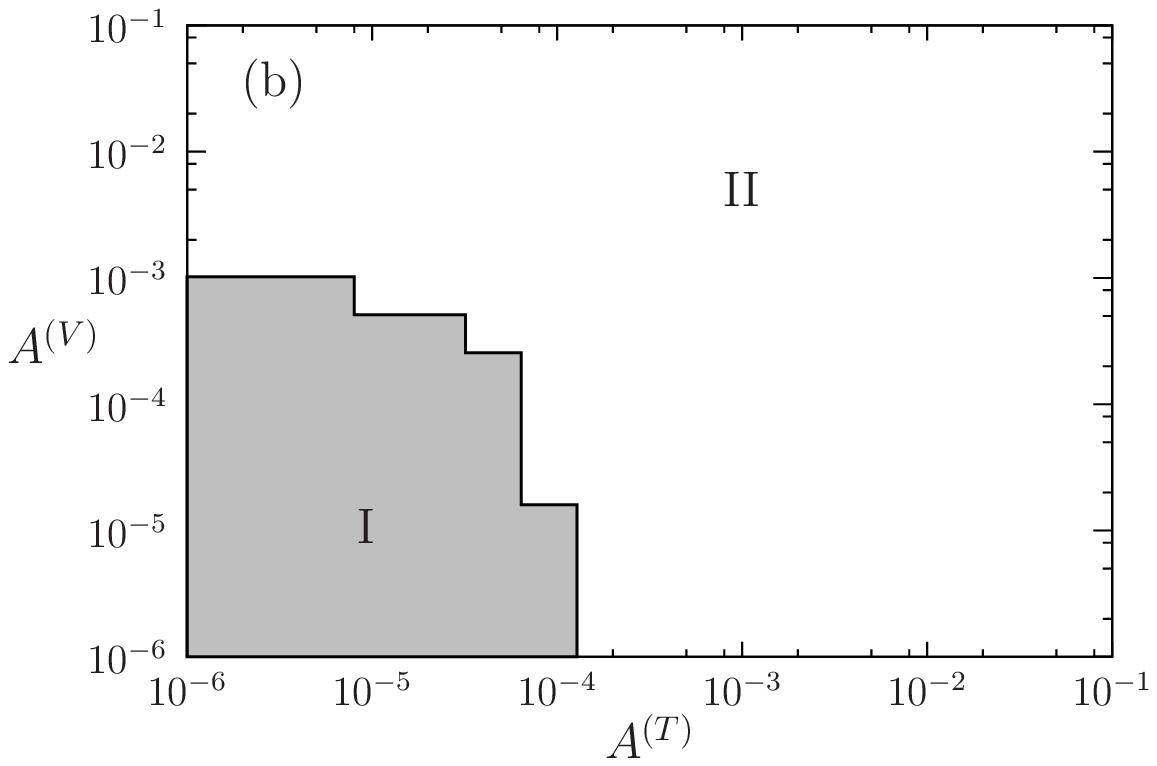} \\
    \includegraphics[width=0.4\textwidth]{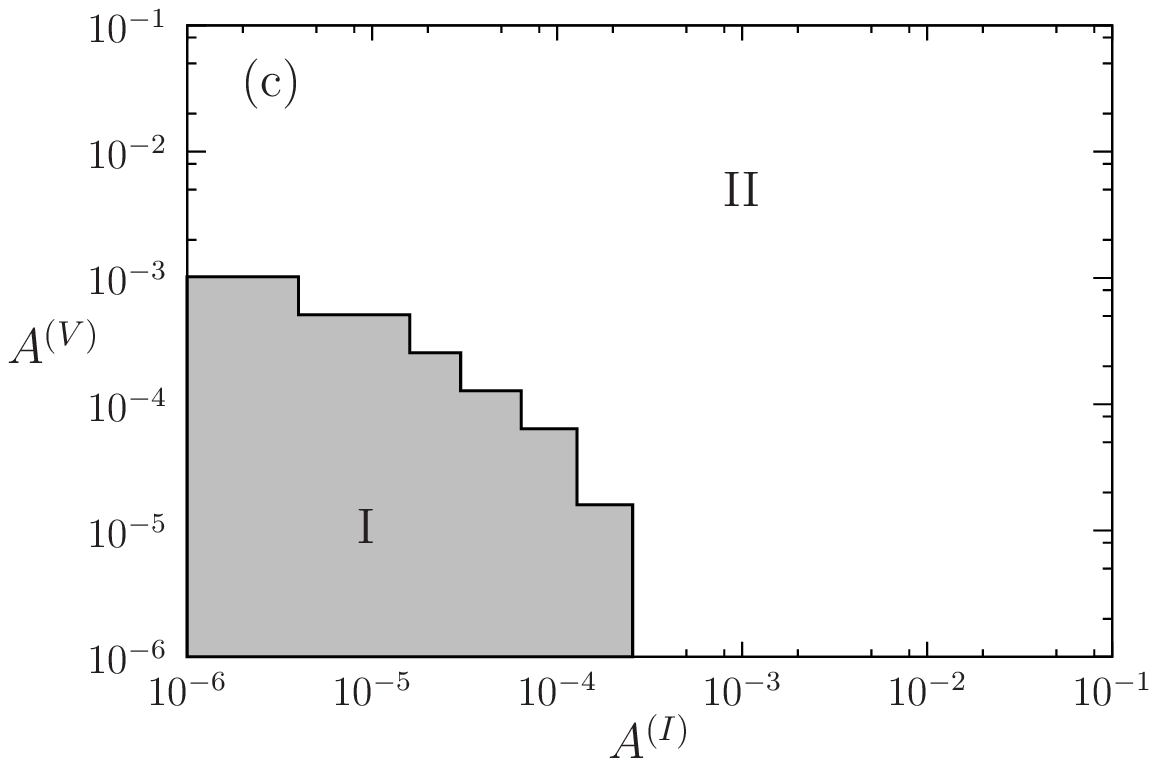} \\
  \caption{ Phase space analysis of $A$s parameters of the $\gamma$
  matrices. 
  In plot (a) we fix $A^{(V)}$=$10^{-6}$ and we show
  $A^{(I)}$ vs. $A^{(T)}$. 
  In plot (b) we fix $A^{(I)}$=$10^{-6}$ and we show
  $A^{(V)}$ vs. $A^{(T)}$.
  In plot (c) we fix $A^{(T)}$=$10^{-6}$ and we show
  $A^{(V)}$ vs. $A^{(I)}$.      
  The other parameters are: number of quasi species $N=25$,
  $\varepsilon_A=10^3$, $B=0$ and $\mu=0$. Region I corresponds to the coexistence of immune system
and viruses (chronic infection) while region II corresponds to the
defeat of viral infection.}
  \label{fig3a}
\end{figure}

The evolution of T cells abundances in a scenario of quasispecies 
is shown in ~\ref{fig3}. Note that when the asymptotic state of our model is given by
a fixed point, the asymptotic distribution is insensitive of the
initial conditions, and the strains corresponding to higher fitness
are more abundant. However, one should consider that this asymptotic
state may be reached after such a long time that it may be outside any
practical scenario of the progression of a disease. The role of
mutations in the transitory regime is quite particular. First of all,
mutations are necessary to populate strains outside the first
inoculum, see also Fig.~\ref{fig4}. Mutations affect also the average
fitness and the width of the quasispecies and the Eigen's error threshold
~\cite{BE2005} is a sort of extremal consequence of this effect. However, in
the presence of coupling among strains, due to competition or to a
global constraints (the $K$ parameter), the specific form of 
mutations does not play a fundamental role, see also
Ref.~\cite{BB1997}. 

It's noteworthy that a naive analogy between evolutionary systems and statistical
mechanics can be obtained by identifying the selection as the
(opposite) of energy and mutation as a sort of temperature
(\cite{T1992,L1987}).

\subsection{Phase diagrams of $\gamma$ matrices}

We focused our attention on the parameter $A$ of the three $\gamma$ matrices 
which represent the major features of the interaction, Fig.~\ref{fig3a}.
The regions I and II in the three diagrams represent chronic infection
(coexistence of immune system and viral infection) and complete
recovery (virus defeat) from the disease, respectively.
For the sake of clarity instead of 3D diagrams, we choose
2D projections, after setting the other dimension to a biologically
meaningful value.

The effect of B cells is shown in Fig.~\ref{fig3a}a : for a very low 
contribution of B cells to the immune response, there is a clear threshold 
effect on both $\gamma^{(T)}$ and $\gamma^{(I)}$. By considering a weak action of T killer cells,  the
threshold effect for $\gamma^{(T)}$ and $\gamma^{(V)}$ (B cells
effect) is somewhat decreased but still present, Fig.~\ref{fig3a}b.
Instead, Fig.~\ref{fig3a}c shows that the combined effects of B and T cells 
are almost additive, despite the parameter $K$ that limits the total 
population of T cells. For example, let's look at diagram (c). 
If we draw a line at
$A^{(I)}=2.56 \cdot 10^{-4}$, that is an extreme value for coexistence
in diagram (a), we can notice that both the two phases are present,
in particular there are values of $A^{(V)}$ leading to the defeat of
the infection.

\subsection{Coinfection dynamics}

\begin{figure}[t]
  \centering
  \includegraphics[width=7cm]{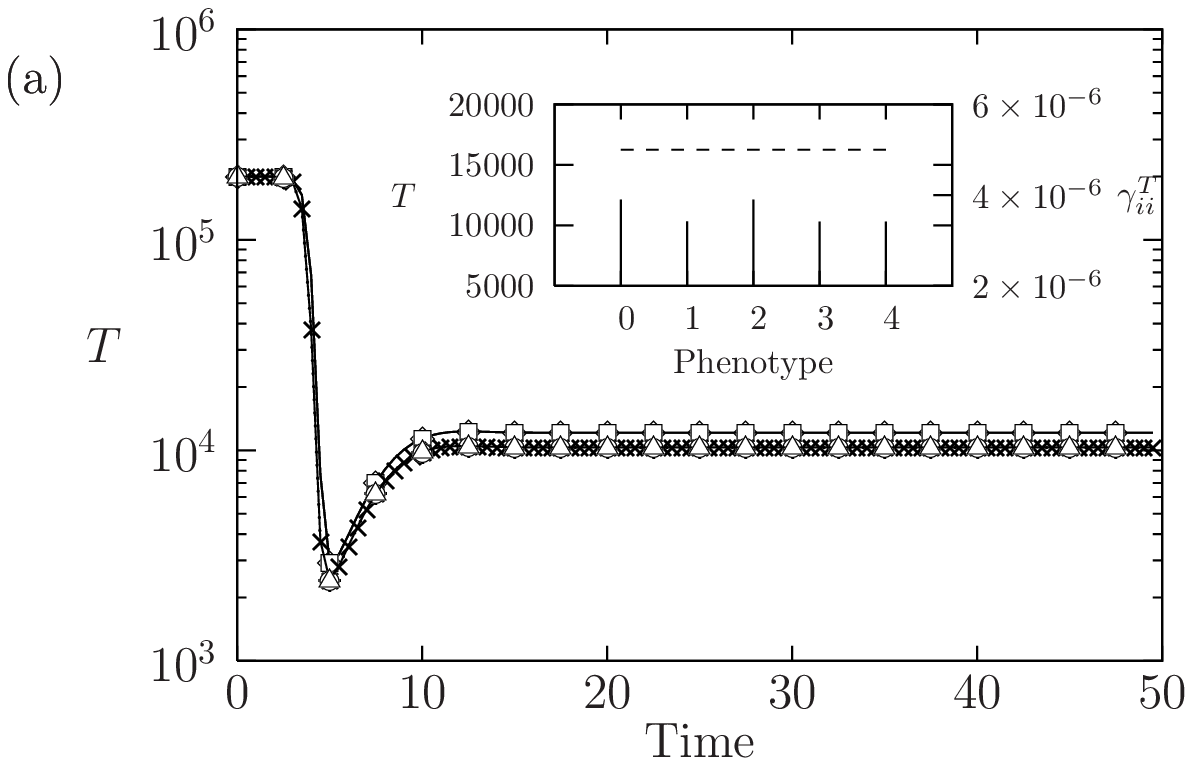} \includegraphics[width=7cm]{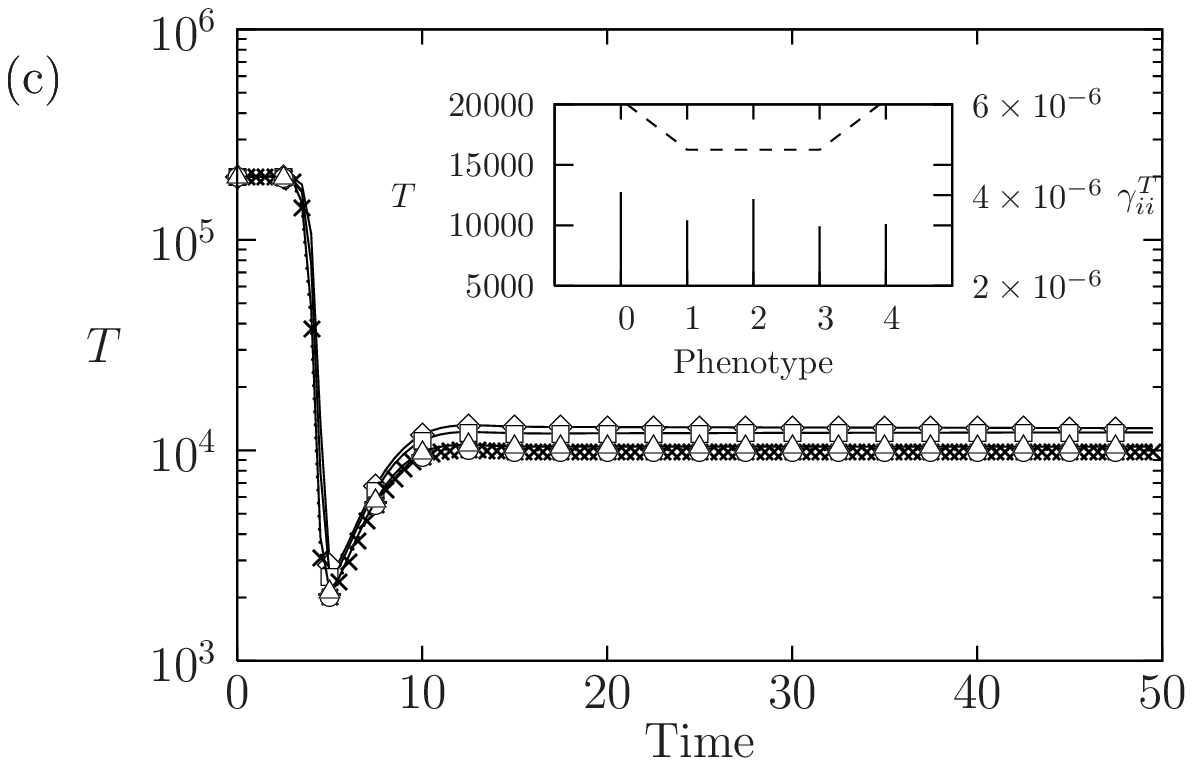} \\
  \includegraphics[width=7cm]{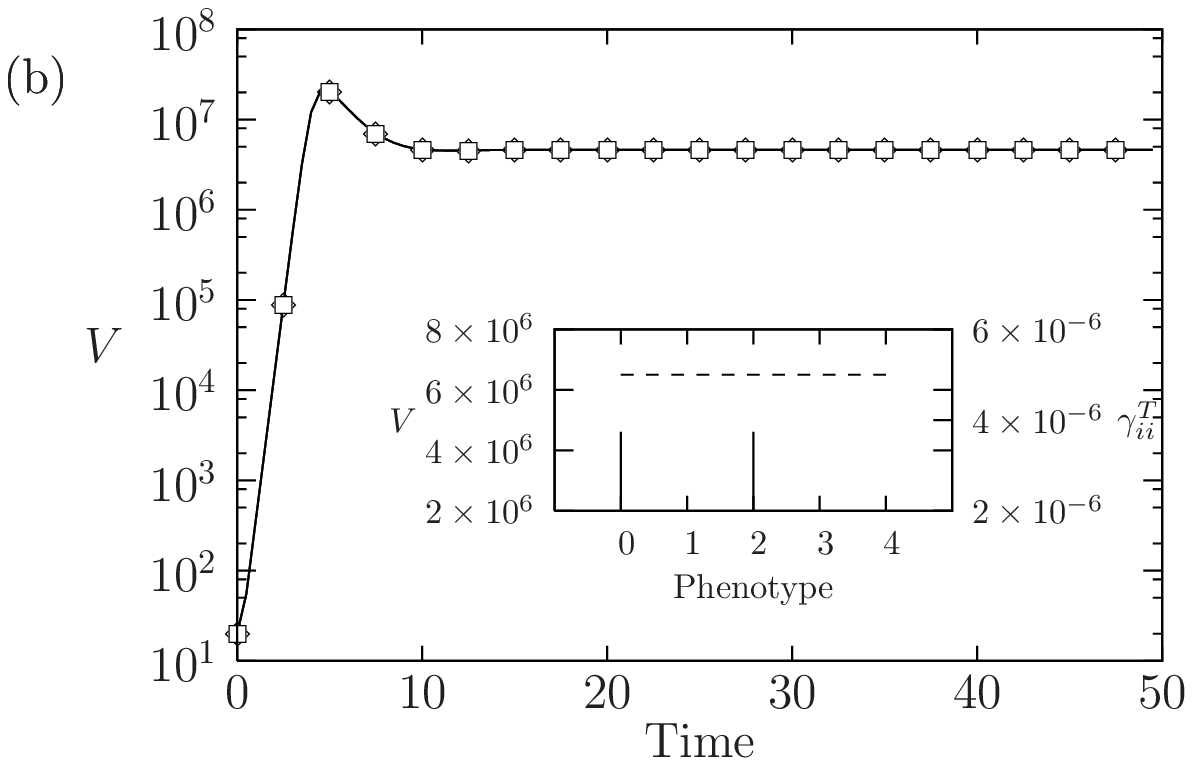} \includegraphics[width=7cm]{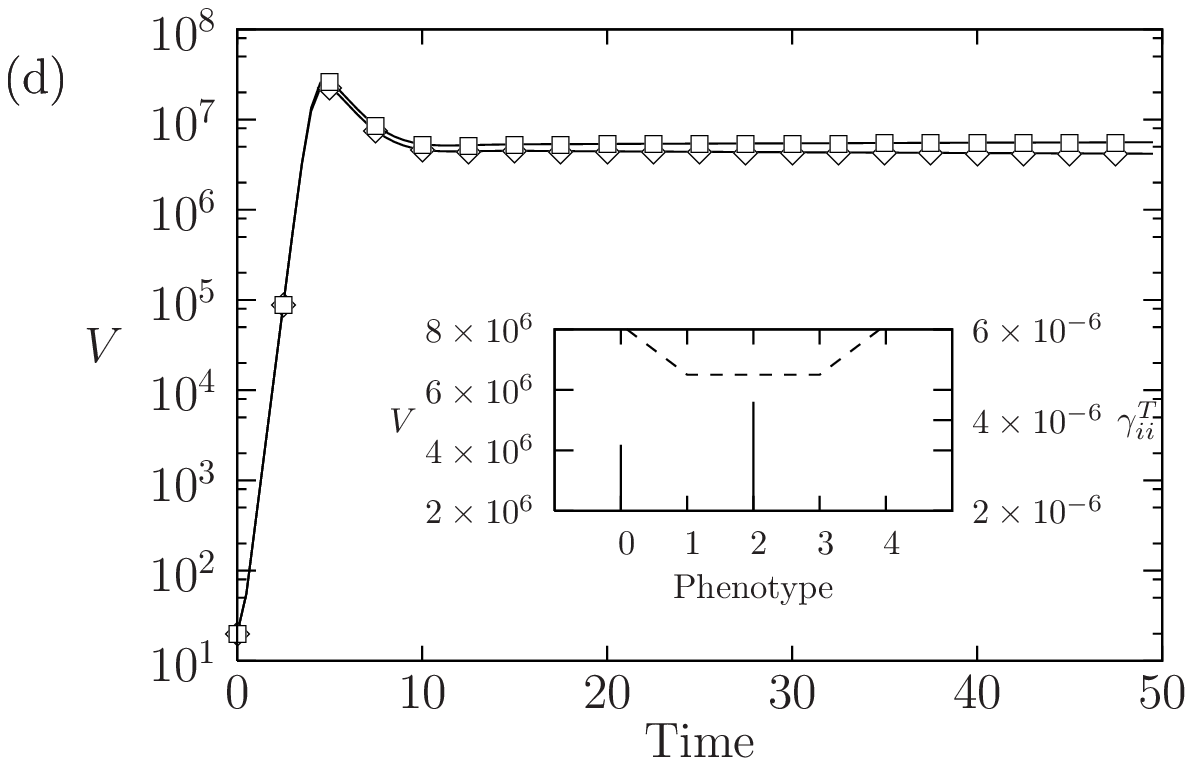} \\
    \caption{Time evolution of T (plots (a) and (c)) and V
    (plots (b) and (d)) populations after an inoculum with 
    two strains (phenotypes 0 and 2), $\mu=0$,  
    $A = 5\cdot 10^{-6}$, $\varepsilon_A=10$, $N=5$. Plots (a) and (b):  $B=0$;
    Plots (c) and (d): $B^{(T)} = 3\cdot 10^{-6}$,
    $\varepsilon_B^{(T)}=10$. In the inset the asymptotic distribution
    of T and V populations and the shape of the diagonal of $\gamma^{(T)}$ are shown. 
    }
  \label{fig4}
\end{figure}

The effect of competitive evolution observed after a single
inoculum of two different HIV strains at time $t=0$ is shown in
Fig.~\ref{fig4}.    
We have considered two different scenarios. The first one
(Fig.~\ref{fig4}a-b) represents the time evolution of T cells
and viruses when the epitopes of two different strains are subjected
to the same interaction strength.
The second scenario (Fig.~\ref{fig4}c-d) considers differences in the
recognition ability of viral antigens by T cells.
For example, variants of the
CCR5 receptors may induce partial resistance against HIV~\cite{GM2003}. 
De Boer and Perelson have shown~\cite{DP1995} that the phase space of the antigen-T cell
recognition is not homogeneous but it is patched with areas of
strong immune response and areas of lack of immune response. 
This difference in the strains targeted by the T cells generates changes in the 
fitness of the virus that turn into differences in their abundances. The immune system response is
almost the same in the two cases, and very similar to that obtained
without considering the phenotypic space, Fig.~\ref{fig2}. The asymptotic distribution of viruses reflects the
behavior of the fitness is induced by the stimulation
of the immune response, $\gamma_{ii}^{(T)}$, as in Fig.~\ref{fig3}.

\subsection{Superinfection dynamics}

\begin{figure}[t]
  \centering
  \includegraphics[width=0.6\textwidth]{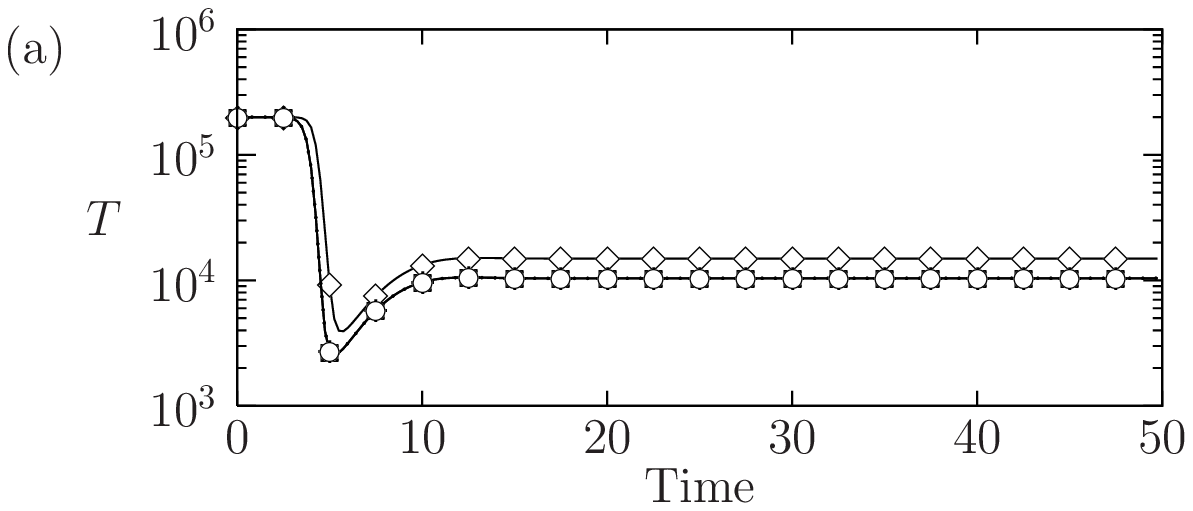} \\
  \includegraphics[width=0.6\textwidth]{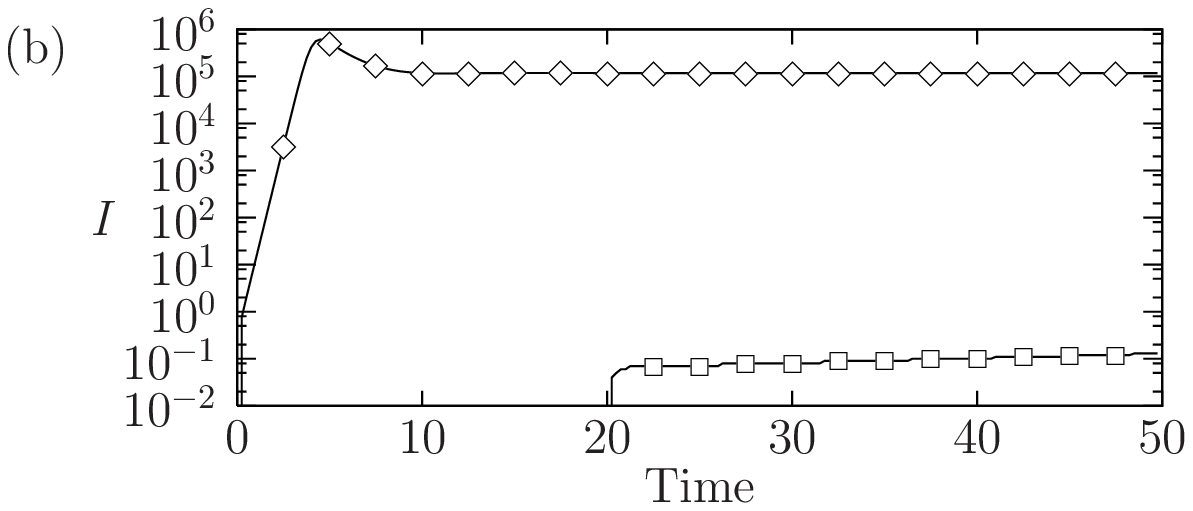} \\
    \includegraphics[width=0.6\textwidth]{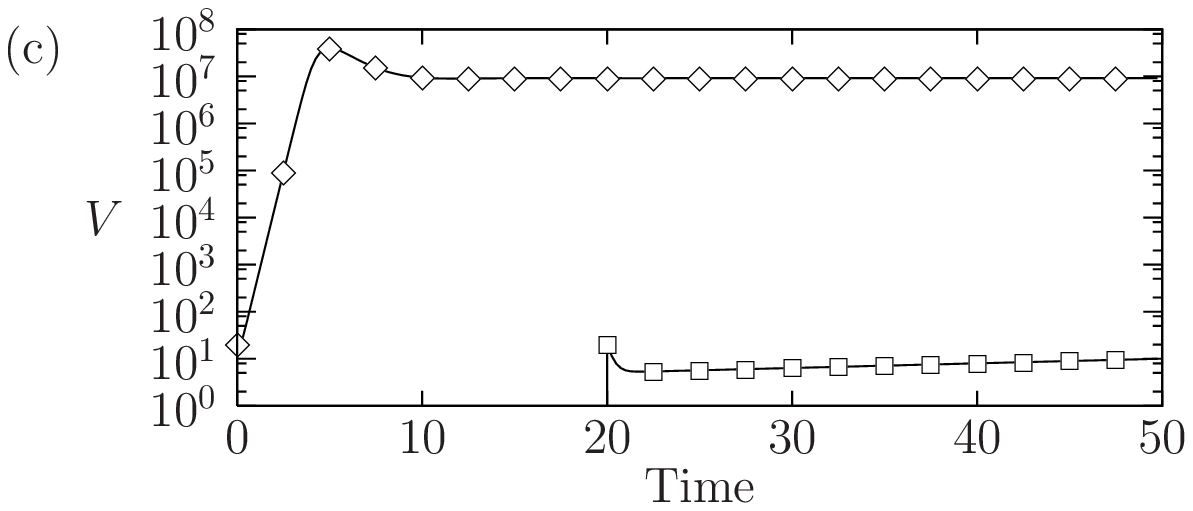} \\
    \caption{Short time behavior of uninfected T cells (a), 
      infected T cells (b) and
      viruses (c) under superinfection. Parameters as in 
Fig.~\ref{fig4}a-b.}
  \label{fig5}
\end{figure}

\begin{figure}[t]
  \centering
  \includegraphics[width=0.6\textwidth]{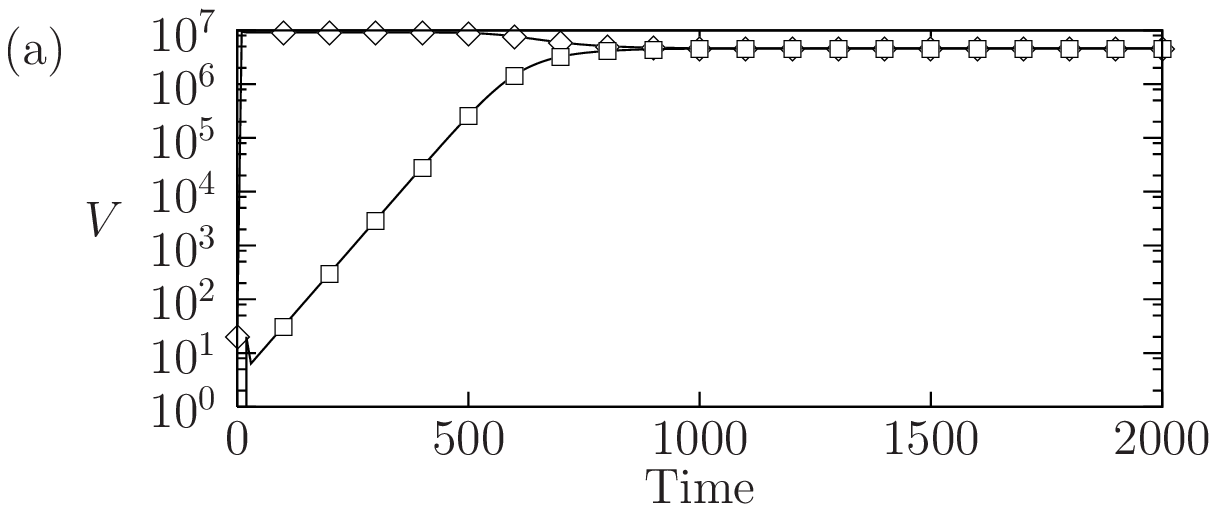} 
  \includegraphics[width=0.6\textwidth]{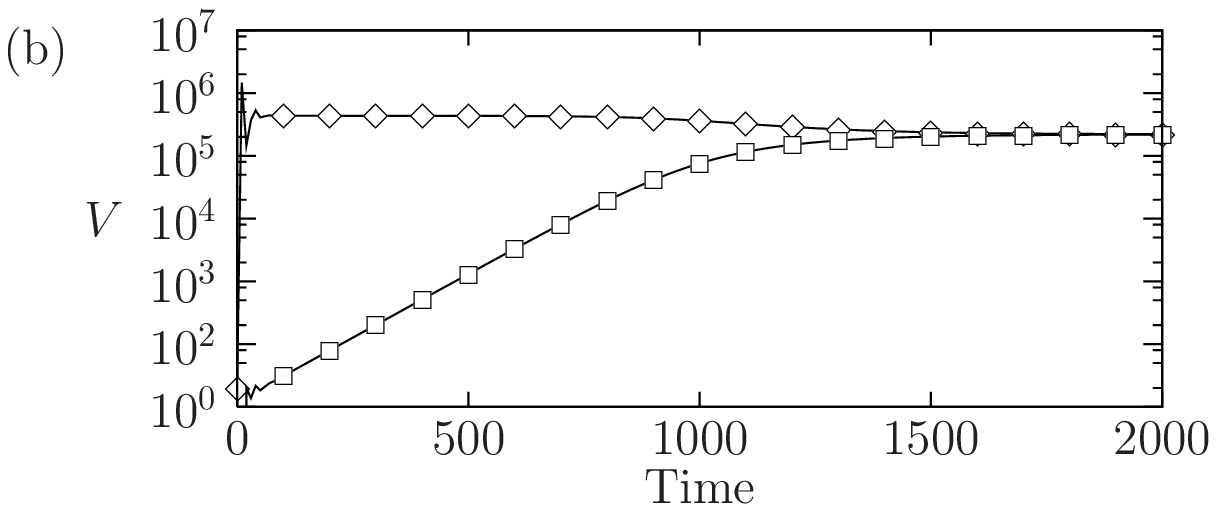} \\
    \caption{Long time behavior of virus abundances after superinfection
    at time $t=20$.  
    (a) healthy immune system ($\lambda=10^5$); (b)
    weak immune system ($\lambda=10^4$). 
    The other parameters used have  the same values 
    of  Fig.~\ref{fig4}a-b.
    Diamonds represent the first
    inoculum and squares the second inoculum.}
  \label{fig6}
\end{figure}

Results of short time and long term viral coevolution after superinfection are 
reported Figs.~\ref{fig5} and~\ref{fig6}, respectively. In both figures the first 
viral inoculum occurs at time $t=0 $ and the second infection at time $t=20$, 
when the immune response to the first inoculum has completed and the virus has 
established a chronic infection. After the second inoculum and a short transient, a
slow mounting of the second viral infection occurs. This low dynamics
continues on a scale of several months (Figure~\ref{fig6}) and 
eventually reaches the same level of the other quasispecies. This 
behavior represents another example of a slow relaxation to a fixed-point 
equilibrium. With the progression of the disease, when the immune system
is compromised (low number of T cells \emph{i.e.} low $\lambda$), 
a second inoculum strain requires a very long time to reach the same abundances 
of the first strain (Fig.~\ref{fig6}b).
 
Experimental evidences~\cite{JD2000} show that
the probability of observing new fitter recombinant strains 
increases with the number of already coevolving strains. This effect
has not been yet incorporated into our model.

\subsection{Coevolution and speciation}

\begin{figure}[t]
  \centering
  \includegraphics[width=6.2cm]{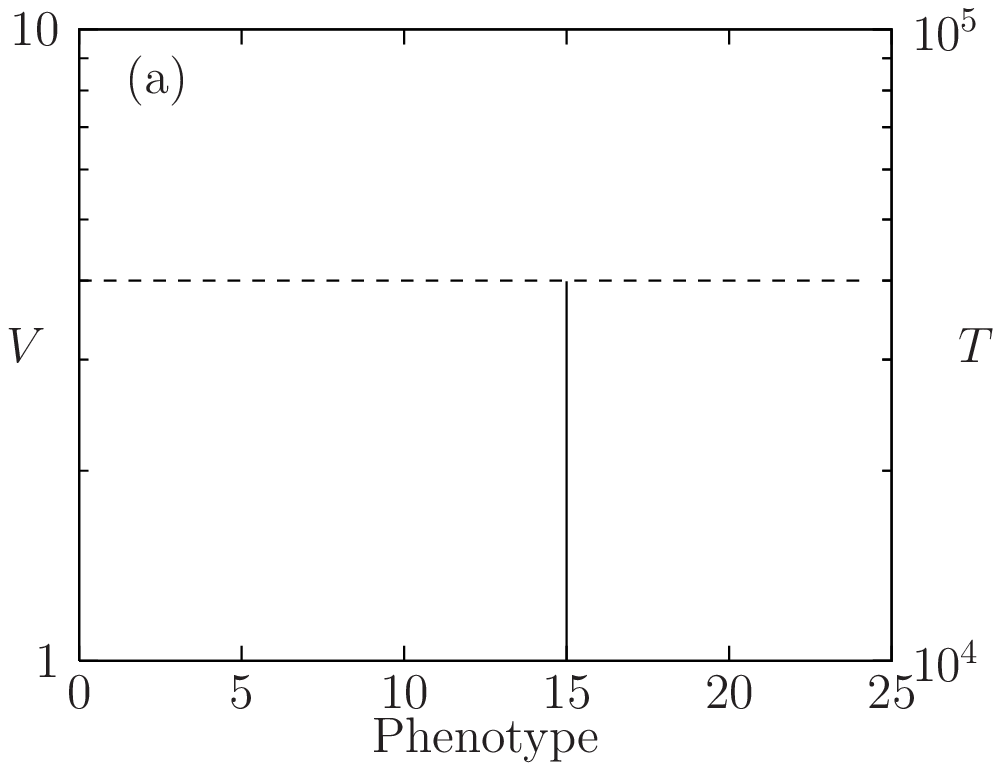} \includegraphics[width=7cm]{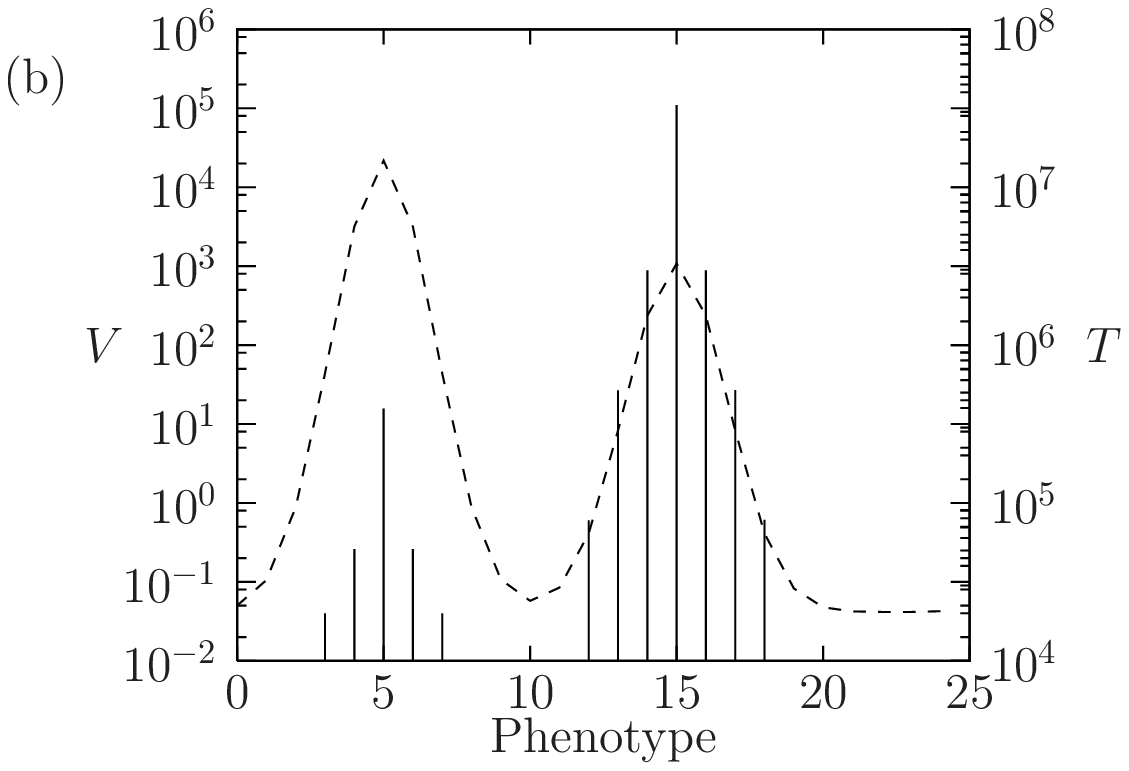} \\
  \includegraphics[width=7cm]{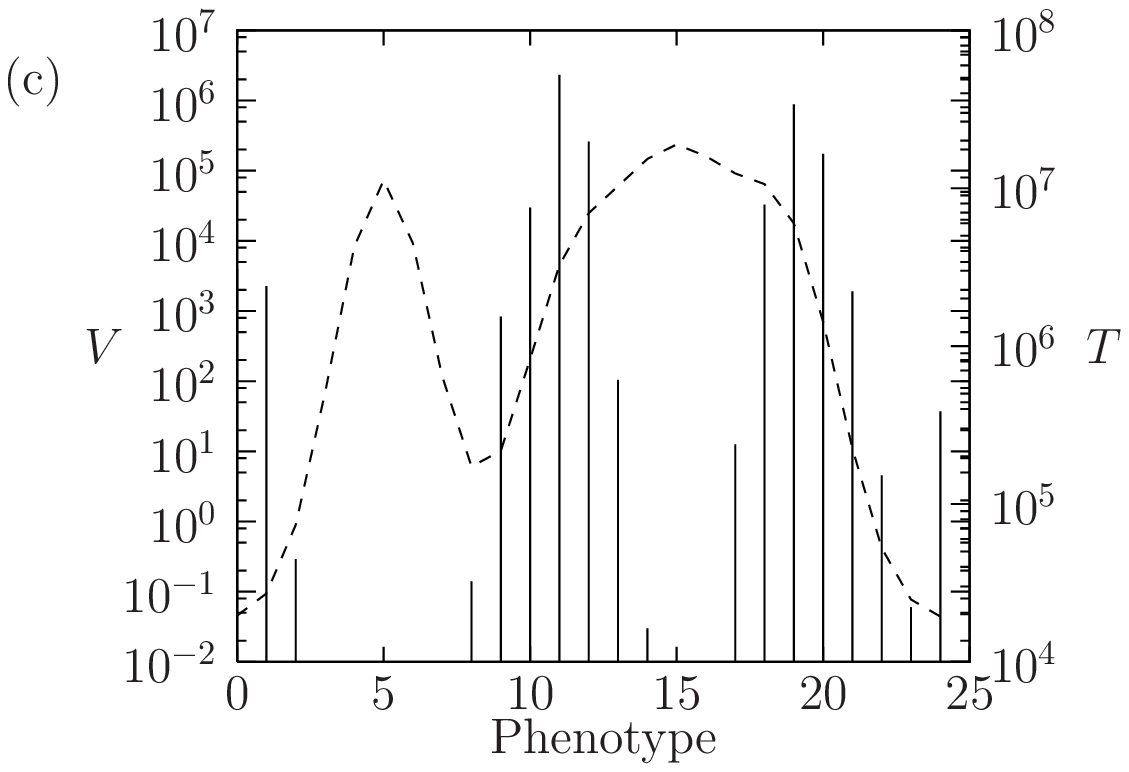} \includegraphics[width=7cm]{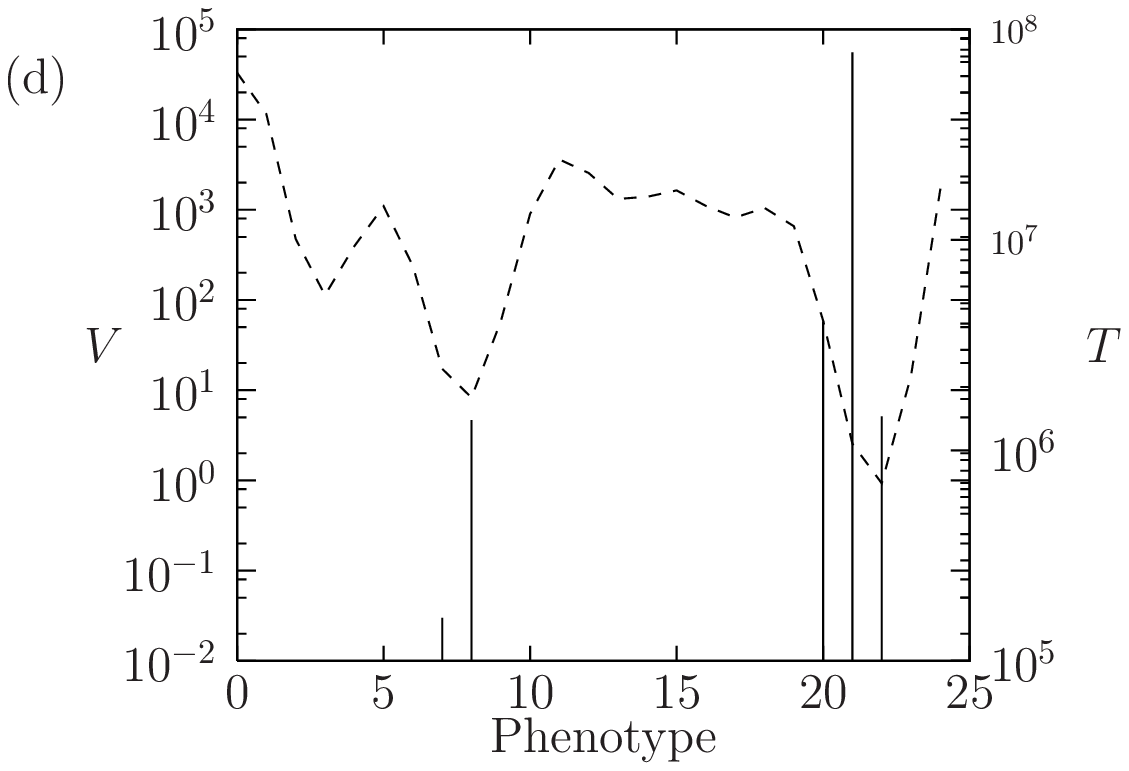} \\
    \caption{Speciation of virus quasispecies and uninfected T cells dynamics
after competitive superinfection at four different times: $t=0$
(a), $t=4.5$ (b), $t=5.25$ (c) and $t=5.75$ (d).
Virus strain 15 is present at time $t=0$, while strain 5 is
inoculated at time $t=1$. Mutation rate $\mu=10^{-4}$ and
non-uniform interaction strength as in Fig.~\ref{fig4}c-d. The dashed
line represents the abundances of T cells targeting
each viral phenotype.}
  \label{fig7}
\end{figure}

\begin{figure}[t]
  \centering
  \includegraphics[width=0.6\textwidth]{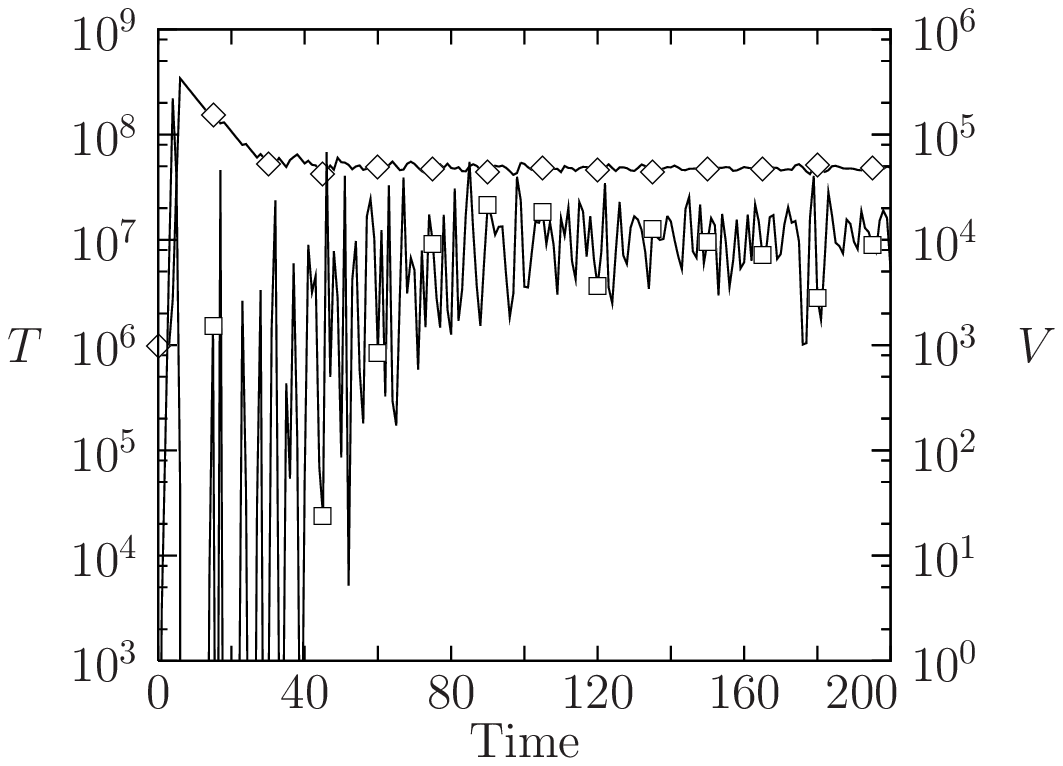}
    \caption{Temporal evolution of speciation of virus quasispecies and uninfected T cells dynamics
    after superinfection; diamonds represent uninfected T cells,
    squares represent viruses. 
    Virus strain with phenotype~15 is present at time $t=0$ and the strain
    with phenotype~5 is inoculated at time $t=1$. 
    We set $A^{(T)} = 5\cdot 10^{-5}$, $A^{(I,V)} = 5\cdot 10^{-6}$, $\varepsilon_A=10$, $B^{(T)} = 3\cdot
    10^{-6}$, $\varepsilon_B^{(T)}=10$, with mutation rate $\mu=10^{-4}$.}
  \label{fig8}
\end{figure}

Finally, we have studied virus quasispecies formation in
more details. Figure~\ref{fig7}a shows the initial  
inoculum at phenotype 15 in a space of 25, followed by a delayed
inoculum at phenotype~5 at time $t=1$. The
non-uniform interaction strength would favor the central phenotypes.
However, since in this simulation the immune system does not discriminate
among similar phenotypes (modeled by $\varepsilon_A^{(T)}=10$),
there is an induced competition among neighboring strains. This
competition induces a separation
of the original quasi-species into two clusters (quasi-speciation), 
Fig.~\ref{fig7}b. However, as shown by Figs.~\ref{fig7}c-d
the immune system response continues to change in time, resulting into
a complex  coevolution with viral populations. 
Fig.~\ref{fig8} shows
the typical irregular evolution of a speciation dynamics.

\section{Discussion}
We presented a model of the within-patience persistence of HIV quasispecies. 
We have first extended the Perelson's standard model~\cite{PH1996} to
incorporate B cells response.
The B cells, once activated by specific T cells, can only
act as a predator, and not being directly targeted as a prey by
HIV. We found that this role represents a non negligible
contribution to the immune response in all cases where a virus or
bacterium is targeting (and being targeted by) T cells.
Interestingly, T cells have a role in both innate and adaptive
immune responses while B cells only in the adaptive system. It is
known that the innate immune response is present also in insects
while the adaptive is more recent, being present only in
vertebrate. Thus, B cells may have appeared also to fight back
viruses targeting specifically at T cells.

Recent works have shown that HIV quasispecies may compete
\cite{CD2000} and that persistence of the initial or ancestor
quasispecies is a good indicator for disease progression
\cite{BL2005}. 

Our model shows that the time evolution of the
competition between quasispecies is slow and has time scales of
several months. This provides a hint of why standard viral dynamics models,
which ignore multiple infections, are effective in describing
viral load evolution in HIV-infected individuals. 

Burch and Chao~\cite{BC2000} have stressed that the evolution
of an RNA virus is determined by its mutational neighborhood.
As the phenotype divergence among viral strains arises from differences 
in selection pressure, these differences may
lead, for instance, to a higher infection rate.
Since the competition is through the immune system response and given
that the phase space of antigen recognition is not homogeneously
covered~\cite{DSP1992}, the HIV high mutation rate allows the
quasispecies to find regions with weak immune response.
This competition may lead to speciation of viral
strains.

The introduction or modulation of a quasispecies may be used in therapy against an
already present aggressive strain. This would be particularly
effective during full AIDS stage when virus burden is particularly
high and more stringent conditions for competition.
It is noteworthy that Schnell and colleagues~\cite{SR1997} have
constructed a recombinant vesicular stomatitis virus that although
unable to infect normal cells, infected and
killed cells that were first infected with HIV causing a rapid
cytopathic infection. The authors showed that the introduction of
this engineered virus can achieve HIV load reduction of 92\% and
recovery of host cells to 17\% of their normal levels (see also
the mathematical model in Ref.~\cite{RG2003}).

Our model represents a general framework to investigate several
aspects of the evolution of HIV infections, for
example intermittency or switching dominance of strains and the
arising of new dominant strains during different phases of
therapy; how superinfection will evolve in case of replacement of
drug-resistant virus with a drug-sensitive virus and acquisition
of highly divergent viruses of different strains. It is also useful 
to investigate
whether antiviral treatment may increase susceptibility to
superinfection by decreasing antigen load. Different drug
treatments can alter the population of quasispecies. Quasispecies
may be the key for both understanding the emerging infectious
diseases and has implications for transmission, public health
counselling, treatment and vaccine development.



\end{document}